\documentclass[twocolumn,msmath,superscriptaddress,amssymb,showpacs]{revtex4}

\usepackage{color} 
\usepackage{graphicx}
\usepackage{epstopdf}
\usepackage{bm}
\usepackage{mathtools}
\usepackage{amsmath}
\usepackage{textcomp}

\begin{document}

\DeclareGraphicsExtensions{.eps, .png, .jpg}
\bibliographystyle{prsty}

\title {Terahertz tuning of Dirac plasmons in Bi$_2$Se$_3$ Topological Insulator}

\author{P. Di Pietro}
\affiliation{Elettra - Sincrotrone Trieste S.C.p.A., S.S. 14 km - 163,5 in Area Science Park, I-34149 Basovizza, Trieste, Italy} 
\author{N. Adhlakha}
\affiliation{Elettra - Sincrotrone Trieste S.C.p.A., S.S. 14 km - 163,5 in Area Science Park, I-34149 Basovizza, Trieste, Italy} 
\author{F.  Piccirilli}
\affiliation{CNR-IOM, Area Science Park, I-34012 Trieste, Italy}
\author{A. Di Gaspare} 
\affiliation {NEST, CNR—NANO and Scuola Normale Superiore, Piazza San Silvestro 12, 56127 Pisa, Italy}
\author{J. Moon}
\affiliation{ Department of Physics and Astronomy Rutgers, The State University of New Jersey 136 Frelinghuysen Road Piscataway, NJ 08854-8019 USA}
\author{S. Oh}
\affiliation{ Department of Physics and Astronomy Rutgers, The State University of New Jersey 136 Frelinghuysen Road Piscataway, NJ 08854-8019 USA}
\author{S. Di Mitri}
\affiliation{Elettra - Sincrotrone Trieste S.C.p.A., S.S. 14 km - 163,5 in Area Science Park, I-34149 Basovizza, Trieste, Italy}  
\author{S. Spampinati}
\affiliation{Elettra - Sincrotrone Trieste S.C.p.A., S.S. 14 km - 163,5 in Area Science Park, I-34149 Basovizza, Trieste, Italy}   
\author{A. Perucchi}
\affiliation{Elettra - Sincrotrone Trieste S.C.p.A., S.S. 14 km - 163,5 in Area Science Park, I-34149 Basovizza, Trieste, Italy} 
\author{S. Lupi}
\affiliation{CNR-IOM and Dipartimento di Fisica, Sapienza Universit\`a di Roma,  P.le Aldo Moro 2, I-00185 Roma, Italy} 
\date{\today}

\begin{abstract}
Light can be strongly confined in sub-wavelength spatial regions through the interaction with plasmons, the collective electronic modes appearing in metals and semiconductors. This confinement, which is particularly important in the terahertz spectral region, amplifies light-matter interaction and provides a powerful mechanism for efficiently generating non-linear optical phenomena. These effects are particularly relevant in Dirac materials like graphene and Topological Insulators, where massless fermions show a naturally non-linear optical behavior in the terahertz range. The strong interaction scenario has been considered so far from the point of view of light. In this paper, we investigate instead the effect of strong interaction on the plasmon itself. In particular, we will show that Dirac plasmons in Bi$_2$Se$_3$ Topological Insulator are strongly renormalized when excited by high-intensity terahertz radiation by displaying a huge red-shift down to 60\% of its characteristic frequency. This opens the road towards tunable terahertz non-linear optical devices based on Topological Insulators. 
\end{abstract}
\pacs{}
\maketitle

Non-linear phenomena play a fundamental role in modern photonics, by enabling many optical functionalities like ultrashort pulse generation and shaping, sum-and difference-frequency processes and ultrafast switching. These effects usually happen in a host material and can be described in terms of an effective photon-photon interaction governed by the material's properties. However, most of non-linear optical effects are extremely weak and have been observed only in the visible/near-infrared spectral regions. In order to enhance the local electromagnetic field and extending non-linear phenomena towards the techonological important terahertz (THz) spectral range, many strategies have been proposed. These proceed along two main directions: The investigation of novel non-linear materials, for instance graphene \cite{mics15,hafez18}, Topological Insulators (TI) \cite{giorgianni17} and 3D Dirac systems \cite{cdas} whose low-energy electrodynamics is characterized by massless Dirac fermions having an intrinsically non-linear THz response \cite{mikhailov17a, mikhailov17b}, and material engineering through cutting edge nanotechnologies \cite{liu12}. A strong non-linear THz response can be obtained by combining these two strategies. 

Topological Insulators are quantum materials exhibiting an insulating electronic gap in the bulk, whose opening is due to strong spin-orbit interaction, and gapless surface states at their interfaces \cite{Moore}. These surface states are metallic, characterized by a Dirac dispersion, showing a chiral spin texture \cite{topo, topo2} and protected from back-scattering by the time-reversal symmetry.  Furthermore, TI surface states spontaneously provide a 2D Dirac-fermion system, segregated from the bulk material without the need of physically implementing an atomic monolayer.
 
The optical properties of TIs are characterized by a free-particle Drude term in the THz region superimposed to a phonon absorption around 2 THz. The Drude term is mainly due to Dirac carriers \cite{AutoreCM} which also substain electronic collective excitations, \emph{i.e.} Dirac plasmons \cite{dipietro13, AutoreCM}.

Plasmon excitations are hybrid modes of charge and light existing at the interface between two media, for instance a dielectric and a metal. They provide the unique possibility to store in sub-wavelength spatial regions the information encoded in a light beam, a property which is of the highest interest for nano-electronics applications and already exploited in nano-(bio) sensors for the detection of vanishingly small quantities of a given analyte \cite{toma15, Odeta1}. The subwavelength confinement implies that the electromagnetic energy associated with the light field is highly concentrated in so-called hot-spots, where the interaction of light and matter is strongly enhanced thus allowing for ultrafast processing of optical signals down to the femtosecond timescale.
This strong interaction scenario has been considered so far mainly from the point of view of light, with the plasmon hot-spots providing an efficient transduction mechanism to induce high-order light-mixing processes and enabling new optical functionalities \cite{kauranen12}. The study of the non-linear effects on the plasmon itself, like changes on its characteristic excitation frequency due to the strong (non-linear) interaction with light, have been much less addressed so far from an experimental viewpoint. 

\begin{figure*}[t]
\begin{center}

\leavevmode
\includegraphics [width=15cm]{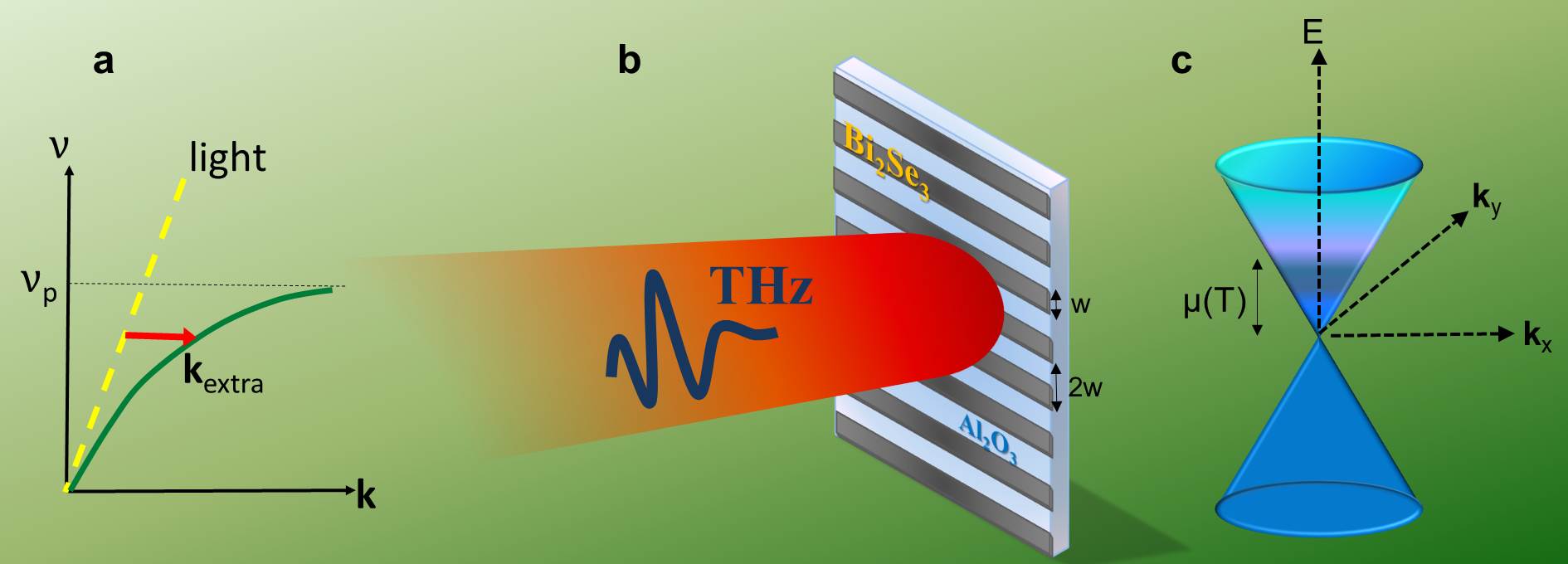}  
\end{center}
\caption{{\bf Plasmons in a Topological Insulator.} Plasmons are collective oscillations of electrons that can be directly excited by electromagnetic radiation in the presence of an extra momentum $k_{extra}$. This is achieved in the present experiment (a,b), through ribbon arrays (width $w$, period 2$w$, and $k_{extra}=\pi/w$), fabricated onto the surface of Topological Insulator Bi$_2$Se$_3$ films (a) with linearly-dispersive electronic bands (c). Plasmons are excited after illumination with sub-ps, half-cycle THz pulses produced at the FERMI free-electron-laser \cite{perucchi13}.}
\label{Fig1}
\end{figure*}

In this paper, we fill this gap by studying the plasmonic response in the terahertz region in Bi$_2$Se$_3$ TI thin films when Dirac plasmons are excited by a strong THz electric field in the MV/cm range. Bi$_2$Se$_3$ TI Molecular Beam Epitaxy (MBE)-based thin films are particularly suitable for optical measurement. Indeed, through the MBE growing technique, one minimizes defects and out-of-stoichiometry, strongly reducing extrinsic optical absorption from the bulk and obtaining a nearly pure Dirac response from surface \cite{oh, giorgianni17, AutoreCM}. A THz plasmonic excitation can be achieved if the TI surface is patterned, for instance in the form of parallel ribbons, thereby providing to THz light the necessary momentum to match, in the absorption process, the plasmon's dispersion relation (see Fig. \ref{Fig1}). Dirac based THz plasmons have been indeed observed in the past few years, both in graphene \cite{ju11} and TIs \cite{dipietro13,autore1,autore2,autore3,lederman,politano}.
We show that Dirac plasmon resonances in Bi$_2$Se$_3$ nanoribbon films are strongly renormalized upon intense THz illumination, by displaying a red-shift down to 60\% of their linear value measured at low $E_{THz}$ field. Our observation proves that TI nanoribbons provide a viable way towards a {\it light-only} active ultrafast plasmon manipulation and tunable sensors.

\section*{Results}

\begin{figure*}[t]
\begin{center}
\leavevmode
\includegraphics [width=12cm]{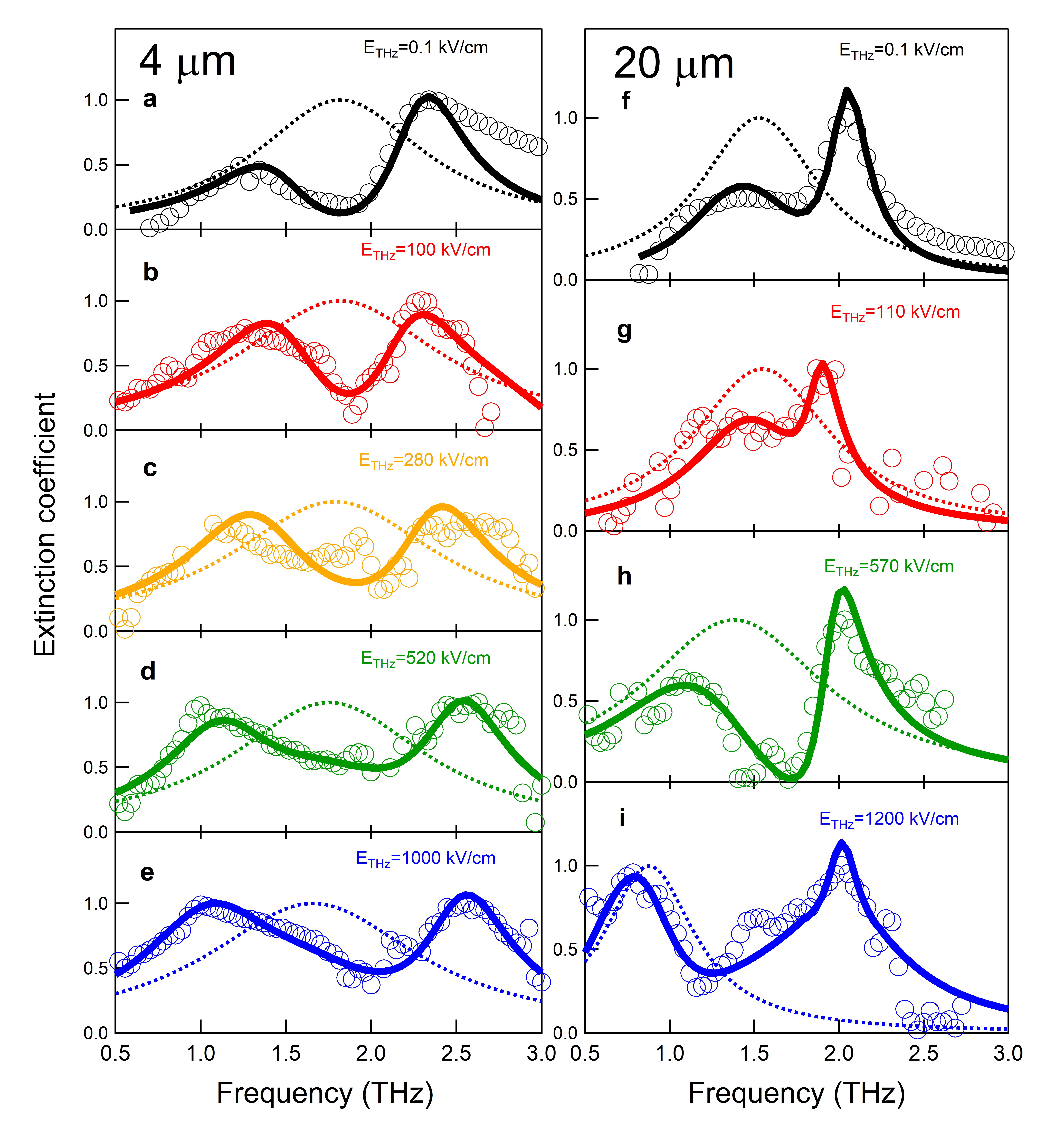}  
\end{center}
\caption{{\bf Fluence-dependent THz spectra}. Panels (a)-(e) display the THz extinction coefficient E($\nu$) of the $w$=4 $\mu$m ribbon array patterned film for different THz electric field values. THz light is perpendicularly polarized to the ribbons. E($\nu$) shows a plasmon-phonon coupled spectrum (see main text). Empty circles are the experimental data points, while the full line is a fit according to Equation (\ref{fano}). The same is shown in panels (f)-(i) for the $w$=20 $\mu$m patterned film. The dashed lines in all panels represent the bare plasmon absorption, as calculated from Equation (\ref{fano}).}
\label{Fig2}
\end{figure*}

We have prepared two different patterned films of Bi$_2$Se$_3$ in form of parallel ribbons with a period $2w$ and widths $w$= 4 and 20 $\mu$m (filling factor 1/2). The THz transmittance T($\nu$) in the linear regime were measured with the synchrotron-based FTIR set-up of the SISSI beamline \cite{lupi1} at the Elettra light source, providing THz pulses at 500 MHz repetition rate with an associated electric field of about 0.1 kV/cm. In order to investigate a possible THz induced non-linear plasmon behavior, we have employed the TeraFERMI \cite{perucchi13, dipietro17} source at FERMI@Elettra Free-Electron Laser, providing high-power broadband sub-ps THz pulses, coupled with a home-made step-scan Michelson interferometer. The extinction coefficient E($\nu$)=1-T($\nu$), as extracted from transmittance T($\nu$) measurements with light polarized perpendicular to the ribbons, is reported in Fig. \ref{Fig2} for the two films.

In the linear regime ($i.e.$ for $E_{THz}$=0.1 kV/cm) one identifies a two hump structure (see Fig. \ref{Fig2}) in the extinction spectrum of the Bi$_2$Se$_3$ nanoribbons patterned films. As already shown in Ref. \onlinecite{dipietro13}, this peculiar absorption feature can be attributed to the THz plasmon resonance which is activated by the patterned structure, coupled $via$ a Fano interaction, with the $\alpha$ phonon of Bi$_2$Se$_3$ at nearly 2 THz. 

In the long wavelength limit  $k\rightarrow0$ (the limit valid in our case) the Dirac plasmon dispersion can be written as \cite{Stauber}:

\begin{equation}
\nu_p=\sqrt{\frac{\pi}{w}}(\frac{e^2}{4\pi\epsilon_0\bar{\epsilon}\hbar}v_F\sqrt{2\pi n g_sg_v})^{1/2}, \label{plasmon}
\end{equation}
where $\bar{\epsilon}$ is an effective dielectric constant ($i. e.$ the average between the vacuum dielectric function and that of the sapphire substrate), $v_F$ the Fermi velocity, $n$ the 2D electron density, and $g_s$=$g_v$=1 are the spin and valley degeneracies, respectively.  

The plasmon and $\alpha$ phonon are mutually interacting $via$ a Fano interference \cite{dipietro13}, therefore the corresponding extinction coefficient can be described through the following Equation obtained by Giannini et al. \cite{giannini}:

\begin{equation}
E(\nu')=\frac{(\nu'+q(\nu'))^2}{\nu'^2+1}.\frac{g^2}{1+(\frac{\nu-\nu_p}{\Gamma_p/2})^2}, \label{fano}
\end{equation}
where $\nu_p$ is the bare plasmon frequency and $\nu_{ph}$ is the bare phonon frequency. 
The expressions for the renormalized frequency $\nu'$ and for the Fano parameter $q$,  that is the ratio between the probability amplitude of exciting a discrete state (phonon) or a continuum or quasi-continuum state (plasmon), are reported in the Supplemental Material (SM). $\Gamma_P$ and $\nu_P$ are the plasmon bare linewidth and frequency, respectively, and $g$ is its coupling factor with the radiation.

\begin{figure*}[t]
\begin{center}
\leavevmode
\includegraphics [width=15cm]{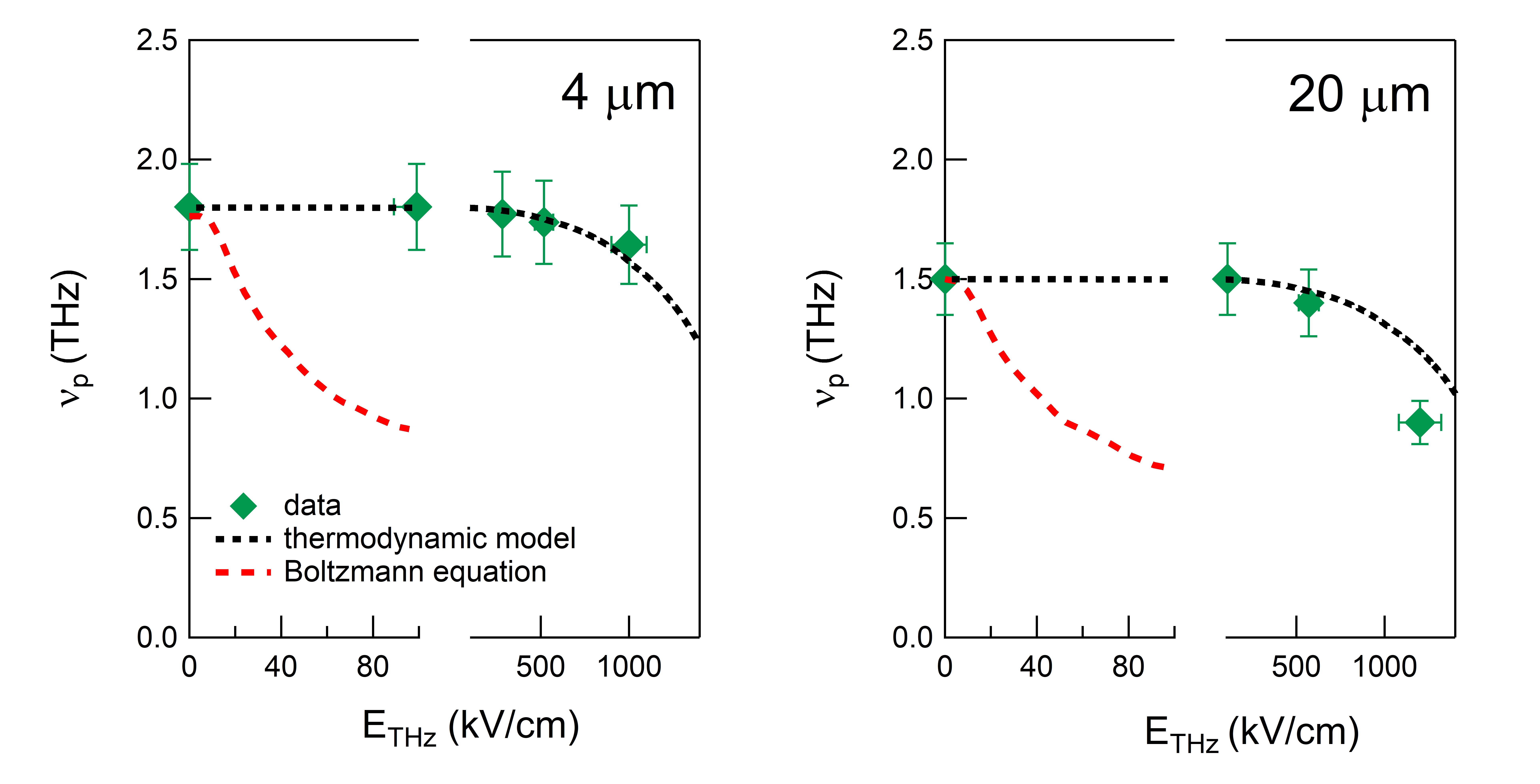}  
\end{center}
\caption{{\bf Plasmon softening.}
Experimental plasmon frequency $\nu_p$ values for both patterned films are shown here as a function of the THz electric field  $E_{THz}$. According to the Boltzmann equation formalism a softening of the plasmon (dashed red line) is theoretically predicted \cite{mikhailov17b}. This model fails in reproducing the experimental data. On the other hand, a thermodynamic model (black dotted line) correctly catches our experimental observations (see main text).}
\label{Fig3}
\end{figure*}

After fitting data from Figure \ref{Fig2}, to Equation (\ref{fano}), we can establish the THz electric field dependence of the bare plasmon frequency $\nu_p$, as reported in Figure \ref{Fig3} (see the Method section for the $E_{THz}$ estimate). With increasing field the bare plasmon frequency of the 4 $\mu$m-ribbon film shifts from 1.8 at 0.1 kV/cm (the linear regime value) to 1.6 THz at the highest field ($\sim $1 MV/cm), $i.e.$ 85 \% of its linear value. Similar results are also found for the film with 20 $\mu$m-ribbon pattern for which the field-dependent bare plasmon frequency changes from 1.5 to 0.9 THz, thus resulting in a plasmon softening of about 60\%. Let us notice that the TeraFERMI spectrum provides THz photons whose energy is well below the bulk gap of Bi$_2$Se$_3$, $i.e.$ nearly 300 meV. This is at variance with optical-pump THz probe results where a hardening of the plasmon frequency has been observed \cite{sim} due to an effective increase of the Dirac carrier population determined by the pumping above the optical gap of $Bi_2$Se$_3$.

\section*{Discussion}
A softening of the plasmon frequency at high THz fields was recently predicted in graphene by several authors \cite {mikhailov17b, cox18}. One theoretical approach to the problem consists in solving the Boltzmann equation of transport in a non-perturbative way \cite{mikhailov17a,mikhailov17b}. The non-linear properties can be completely defined by one dimensionless parameter $\mathcal{F}_{\tau}=\tau eE_{THz}/\hbar k_F$, where $e$ is the electron charge, $E_{THz}$, the THz electric field, $\tau$ the scattering rate of free Dirac electrons, and $k_F$ the Fermi wavevector. $\mathcal{F}_{\tau}$  essentially measures the ratio of the energy transferred by the THz pulse to a single electron between two scattering events with respect to the Fermi energy (see SM). The model predicts a very significant plasmon softening at relatively low fields ($\sim$ 10 kV/cm) which is however not found in our experimental data (Fig. \ref{Fig3}, red dashed line). A probable explanation for this disagreement is related to the low heat capacity associated to the massless Dirac electrons. Indeed, the temperature of the electron bath undergoes a strong enhancement upon intense THz excitation (which is actually not considered in the Boltzmann equation model), thus resulting in a non-linear response dominated by incoherent thermal effects \cite{cox18}.

In order to take into account the local variation of temperature due to the absorption of the THz energy, we adopt a standard Two-Temperature model \cite{schoenlein87, lai14}. Here, the electronic temperature $T_e$ of our Bi$_2$Se$_3$ films can be determined by numerically solving the coupled differential equations:

\begin{equation}
c_e(T_e)\frac{dT_e}{dt}=-G(T_e)\cdot(T_e-T_L)+P(t), \label{DiffEq1} 
\end{equation}
\begin{equation}
c_l(T_e)\frac{dT_L}{dt}=G(T_e)\cdot(T_e-T_L), \label{DiffEq2}
\end{equation}

\noindent where T$_e$ and $T_L$ are the electronic and lattice temperatures respectively, $c_e(T_e)$ is the electronic specific heat, $G(T_e)$ describes the coupling between the electron and phonon subsystems, $c_l(T_e)$ is the lattice specific heat, and  $P(t)$ is the THz istantaneous power absorbed by the film. As detailed in SM, all these parameters are available from literature. As a consequence of the THz pulse absorption, the electronic temperature of Bi$_2$Se$_3$ reaches up to 1500 K for fields of about 1 MV/cm. This  value is significantly lower than the Fermi temperature in Bi$_2$Se$_3$. Let us notice that T$_e$ in Bi$_2$Se$_3$ is also lower than in graphene, where, due to a less effective interaction of electrons with the lattice, electronic temperatures as high as 6000 K are observed under the same illumination conditions \cite{jadidi16}. Similarly to graphene, the rise of $T_e$  follows the pump profile $P(t)$ within few 10's of fs, and relaxes back to the lattice temperature within few ps (see SM). 

The increase of the electronic temperature due to the THz illumination can be traced in a strong modification of the electronic chemical potential and, finally, in the softening of the plasmon frequency. Indeed, in a Dirac system the Sommerfeld expansion of the chemical potential at the first order writes $\mu(T_e)\sim\epsilon_F(1-\frac{\pi^2 k_B^2T_e^2}{6\epsilon_F^2})$ \cite{jadidi16,jadidi19}, which directly mirrors in the temperature dependence of $\nu_p\propto \mu(T_e)^{1/2}$.  

\noindent According to a  pure thermodynamic model, this relationship can be used to find the dependence of $\nu_p$ on $E_{THz}$ which is shown in Fig. \ref{Fig3} (black dotted line).
The predicted plasmon frequencies, are in remarkable agreement with the experimental data, with only minor deviations observed at the highest THz field point for the 20 $\mu$m patterned film. 
Moreover, at least at high field and therefore at high induced temperature, one cannot also exclude a thermal excitation of electrons from the surface into the bulk, and therefore a further softening of plasmon not considered in (\ref{DiffEq1}) and (\ref{DiffEq2}). 
The thermodynamic model has been already used to describe the field-dependent THz non-linear properties of graphene, like its optical conductivity \cite{mics15}, harmonics generation \cite{hafez18}, and pump-probe dynamics in nanoribbons \cite{jadidi16,jadidi19}. 
It is also known that when the electronic temperature increases, the electron-electron scattering rate increases as well. Since the Boltzman kinetic equation does not take into account electron-electron many-body effects \cite{sun18a}, its applicability for high THz pump intensities becomes questionable, while a hydrodynamic regime uniquely characterized by frictional forces between electrons may eventually apply for high enough $T_e$ \cite{sun18b}. In the case of TIs, a purely hydrodynamic regime is probably hampered by the presence of electron-phonon scattering channels which are much more significants than in graphene. However the increase in $T_e$ is still high enough to make the Boltzmann kinetic approach not suitable to the description of our data.

We have shown here, that Dirac plasmons in Bi$_2$Se$_3$ topological insulator nanoribbons can be tuned with THz light, over a wavelength range spanning almost one octave. In the regime explored in this paper, the plasmon softening effects can not be described with a conventional Boltzmann equation model, while a thermodynamic picture, similar to that recently applied to graphene, captures most of our findings. The strong renormalization of plasmon excitation observed here, is extremely promising for the exploitation of Topological Insulators as a platform for the realization of active plasmonic devices, their unique properties in terms of spin-polarized currents representing an additional asset for ultrafast nanospintronics applications.

\section*{Methods}
\small
{\bf Sample Preparation.} 
The high-quality Bi$_2$Se$_3$ thin films were prepared by molecular beam epitaxy using the standard two-step growth method developed at Rutgers University\cite{bansal11,oh}. The 10x10 mm$^2$ Al$_2$O$_3$ substrates were first cleaned by heating to 750 $^\circ$C in an oxygen environment to remove organic surface contamination. The substrates were then cooled to 110 $^\circ$C, and an initial three quintuple layers of Bi$_2$Se$_3$ were deposited. This was followed by heating to 220 $^\circ$C, at which the remainder of the film was deposited to achieve the target thickness. The Se:Bi flux ratio was kept to  10:1 to minimize Se vacancies.

Bi$_2$Se$_3$ ribbons were fabricated by electron-beam lithography and subsequent reactive ion etching. The Bi$_2$Se$_3$ film was spin-coated with a double layer of electron- sensitive resist polymer poly(methyl methacrylate) (PMMA) to a total thickness of 1.4 mm. Ribbon patterns with different $w$ were then written in the resist by electron- beam lithography. To obtain a lithographic pattern with a realignment precision below 10 nm over a sample area suitable for terahertz spectroscopy of 10x10 mm$^2$, we used an electron beam writer equipped with an x-y interferometric stage (Vistec EBPG 5000). The patterned resist served as mask for the removal of Bi$_2$Se$_3$ by reactive ion etching at a low microwave power of 45 W to prevent heating of the resist mask. Sulphur hexafluoride (SF$_6$) was used as the active reagent. The Bi$_2$Se$_3$ film was etched at a rate of 20 nm/min, which was verified by atomic force microscopy after soaking the sample in acetone to remove the PMMA. The in-plane edge quality after reactive ion etching, inspected by atomic force microscopy, closely follows that of the resist polymer mask, that is, with an edge roughness of 20 nm. The vertical profile of the edge forms an angle of $\sim$45$^\circ$ with the substrate plane, because our reactive ion etching process has no preferred etching direction.

{\bf The SISSI synchrotron source.} 
The optical measurement at low electric field ($E_{THz}\sim$ 0.1 kV/cm) were performed at the SISSI synchrotron source of the Elettra Storage ring. The SISSI beamline \cite{lupi1, lupi2} provides a high photon flux ($\sim$ mW/mm$^2$) in the THz range, but relatively low peak power and THz field  due to the high repetition range and moderately long pulses ($\sim 30$ ps). The SISSI beamline is coupled to a Bruker Vertex 70v spectrometer equipped with a Si beamsplitter and a He-cooled bolometer detector. The radiation was polarized either along, or perpendicular to the ribbons by a terahertz polarizer with a degree of polarization of 99.5\%. The extinction coefficient reported in this manuscript is obtained from the film transmittance, defined as the ratio between the intensity transmitted by the thin film and that transmitted by the bare substrate.

{\bf The TeraFERMI high power THz source.} 
TeraFERMI \cite{perucchi13,dipietro17} is the superradiant THz beamline of the FERMI free-electron-laser. TeraFERMI produces THz radiation from a Coherent Transition Radiation source in the form of broadband, sub-ps, half-cycle THz pulses at 50 Hz repetition rate. In the present experiment, the radial polarization pattern of Coherent Transition Radiation was made linear with the help of a THz polarizer oriented either perpendicular or parallel to the ribbons. The peak electric field has been established by using the standard formula\cite{ozaki10}

\begin{equation}
E_{THz}=\sqrt{\frac{\eta_0W}{\pi w_I^2 \Delta t } }
\end{equation}

\noindent where $\eta_0$ is the free space impedance (377 $\Omega$), $W$ is the THz energy, $w_I$ is the intensity beam waist, and $\Delta t$ is the THz  pulse duration. These quantities are established by making use of pyroelectric detectors calibrated with a GENTEC THZ12D powermeter, and a Pyrocam III THz camera. The THz pulse duration is assumed to be identical to that of the relativistic electron bunch ($\Delta t=0.8$ ps) employed for its generation. The electric field $E_{THz}$ was changed by attenuating the beam through the use of THz polarizers with variable relative orientation. The THz spectra have been measured with the help of an home-made step-scan Michelson interferometer equipped with a mylar beamsplitter and a He-cooled bolometer detector.

\section*{Acknowledgements}
JM and SO are supported by the Gordon and Betty Moore Foundation’s EPiQS Initiative (GBMF4418) and National Science Foundation (NSF) grant EFMA-1542798.

\section*{Author contributions}
P.D.P., A.P. and S.L. conceived the study. P.D.P, N.A., and F.P. carried out the experiment. P.D.P. and A.P. analysed the data and carried out theoretical modelling. J.M. and S.O. prepared the films. A.D.G. performed lithographic patterning. S.D.M. and S.S. prepared the electron beam for THz emission. P.D.P., A.P., and S.L. wrote the manuscript with contributions from all the authors.

\end{document}